\documentclass[12pt]{article}
\usepackage[margin=0.95 in]{geometry}
\usepackage{amsmath} 
\usepackage{slashed}
\usepackage{comment} 
\usepackage{dsfont}
\usepackage{amssymb,amsfonts}
\usepackage[all]{xy}
\usepackage{graphicx}
\usepackage{amsmath}
\usepackage{amssymb}
\usepackage{float}
\usepackage{array}
\usepackage{tikz}
\usepackage{mathtools}
\usepackage{mathrsfs} 
\usepackage{cite}
\usepackage{tikz}
\usepackage{array, multirow}  
\allowdisplaybreaks
\numberwithin{equation}{section}
\newcommand{\be}{\begin{equation}}
\newcommand{\ee}{\end{equation}}

 \newcommand{\Tr}{{\text{Tr}}}
\def\bea{\begin{eqnarray}}
\def\eea{\end{eqnarray}}
\numberwithin{equation}{section}
\numberwithin{table}{section}

\usepackage[colorlinks=true, citecolor=black, linkcolor=black, filecolor=black,urlcolor=blue]{hyperref}

\begin{document}

\hypersetup{pageanchor=false}
\begin{titlepage}
\vbox{\halign{#\hfil    \cr}}  
\vspace*{15mm}
\begin{center}
{\Large \bf Thermal Correlators and Currents of 
the ${\cal W}_3$ Algebra}

\vspace*{10mm} 

{\large 
Sujay K. Ashok$^{a,b}$, Sanhita Parihar$^{a,b}$, Tanmoy Sengupta$^{a,b}$,\\ 
Adarsh Sudhakar$^{a,b}$, Roberto Tateo$^{c}$ 
}

\vspace*{8mm}

$^a$The Institute of Mathematical Sciences, \\
		 IV Cross Road, C.I.T. Campus, \\
	 Taramani, Chennai, India 600113

\vspace{.6cm}

$^b$Homi Bhabha National Institute,\\ 
Training School Complex, Anushakti Nagar, \\
Mumbai, India 400094

\vspace{.6cm}

$^c$Dipartimento di Fisica, Università di Torino,\\ 
INFN Sezione di Torino,\\ 
Via P. Giuria 1, 10125, Torino, Italy. 

 \vskip 0.8cm
	{\small
		E-mail:
		\texttt{sashok, sanhitap, tsengupta, adarshsu @imsc.res.in} \\
  \texttt{roberto.tateo@unito.it }
	}
\vspace*{0.8cm}
\end{center}

\begin{abstract} 
Two dimensional conformal field theories with the extended ${\cal W}_3$ symmetry algebra have an infinite number of mutually commuting conserved charges, which are referred to as the quantum Boussinesq charges. In this work we construct local operators whose zero modes are precisely these conserved charges. For this purpose we study the higher spin conformal field theory on the torus and compute thermal correlators involving the stress tensor and the spin-3 current in a higher spin module of the ${\cal W}_3$ algebra. In addition we independently obtain the excited state eigenvalues of the quantum Boussinesq charges within the higher spin module via the ODE/IM correspondence. A judicious combination of these data allows us to derive the local operators, whose integrals are the conserved charges of the integrable hierarchy. 

\end{abstract}

\end{titlepage}

\hypersetup{pageanchor=true}

\setcounter{tocdepth}{2}
\tableofcontents

\section{Introduction}

In this work we continue our study of the quantum integrals of motion of two dimensional systems with extended conformal symmetry generated by the ${\cal W}_3$ algebra \cite{Ashok:2024zmw}. The algebra, first studied in \cite{Zamolodchikov:1985wn}, is generated by a spin two current, which is the energy-momentum tensor $T(u)$, and a spin three current, denoted by $W(u)$. We study the theory on a spatial circle, and work with just the chiral half of the conformal field theory.

As in the Virasoro case \cite{Sasaki:1987mm, Eguchi:1989hs, Kupershmidt:1989bf, Bazhanov:1994ft}, the universal enveloping algebra of the ${\cal W}_3$ algebra is expected to have an infinite dimensional Abelian algebra that reflects the underlying integrability of the theory. The first four of these mutually commuting quantum integrals of motion for the ${\cal W}_3$ case were constructed in \cite{Bazhanov:2001xm}, and, following analogous work in the Virasoro case \cite{Bazhanov:1996dr, Bazhanov:1996aq, Bazhanov:1998dq}, the integrable structure of the higher spin conformal field theory was explored in detail in that reference.

In our previous work \cite{Ashok:2024zmw}, we proposed a systematic method to derive the local operators associated with the higher conserved charges of the integrable hierarchy. 
There are two ingredients that go into our construction of the quantum integrals of motion. The first uses the fact that these quantum integrals of motion are expressed as the zero modes of local operators. By virtue of being a zero mode, the action of the Boussinesq charge on the level sub-spaces of a highest weight module is a well-defined linear operator. The crucial input we use is that the spectra of each of these linear operators (at any excited level) can be obtained via the ODE/IM correspondence \cite{Dorey:1998pt, Bazhanov:1998wj, Masoero:2018rel, Masoero:2019wqf}, according to which the eigenvalues of the quantum integrals of motion are encoded in the WKB periods of certain ordinary differential equations. Using this method, we compute the eigenvalues in the excited states at the first and second levels of the higher spin module. 

The second important ingredient is the evaluation of thermal correlation functions of composite operators of the ${\cal W}_3$ algebra. This is carried out by making use of the Zhu recursion relations \cite{zhu:1990, zhu:1996}. We combine these two diverse ingredients and show that it is possible to systematically derive the higher quantum integrals of motion of the Boussinesq hierarchy. 

The present work is a direct follow-up to our previous paper \cite{Ashok:2024zmw}, and the basic strategy we have used to fix the currents is already spelt out in that reference. However, we were unable to fix the higher currents in that reference as we had worked out the eigenvalues of the quantum Boussinesq charges only at the first excited level of the higher spin module. In this work, we proceed to calculate the eigenvalues at the second excited level and, in addition, determine the two-point functions of the charges with zero modes of multiple local operators to fix the form of the currents. This work, therefore, illustrates, to a large extent, the power of the formalism we use to find the higher currents. From a physics point of view, fixing the form of the currents and calculating the thermal correlators has many potential applications, especially to the generalized Gibbs ensemble \cite{Maloney:2018yrz, Dymarsky:2018lhf} and the Eigenstate Thermalization Hypothesis for conformal field theories \cite{Lashkari:2016vgj, Dymarsky:2019etq}. Finally, the analysis presented in this paper and its potential applications provide a new concrete example of the effectiveness of the ODE/IM correspondence in addressing interesting open problems in two-dimensional conformal field theory. 

This paper is organized as follows. In section \ref{thestrategy}, we give a step-by-step breakdown of the strategy to fix the currents. In section \ref{operatorformalism}, we exhibit the action of various relevant linear operators on the excited states of the higher spin module, and in section \ref{level2eigenvalues}, we summarize how the excited state eigenvalues are calculated. Section \ref{highercurrents} combines the results for the spectrum from the ODE/IM correspondence with the thermal correlators to compute the Boussinesq currents. Some of the technical details associated with the computation are collected in the appendices.

\section{The Strategy}
\label{thestrategy}

In this section we first review some of the relevant results obtained in \cite{Ashok:2024zmw} and describe the particular strategy we shall follow in our analysis. Our main goal is to derive the currents $J_{n+1}$ whose  integral over the spatial circle provides the conserved charges of the quantum Boussinesq hierarchy: 
\be 
{\bf I}_n = \int_0^{1}\, du\, J_{n+1}(u)~.
\ee 
Here $u$ is the periodic coordinate on the spatial circle, with $u\sim u+1$.  The subscript indicates the operator's conformal weight. 
Each current is a linear combination of composite operators of a given conformal weight, and each composite operator is (conformally) normal ordered. Since all currents up to weight\footnote{For the ${\cal W}_3$ algebra, the conserved charges ${\bf I}_{3n}$ are zero, and so we will not consider the corresponding currents $J_{3n+1}$ of weight $3n+1$.} eight were derived in \cite{Ashok:2024zmw},  we will start by looking at the currents of weight nine to illustrate our methodology.

\paragraph{1.} Our starting point will always be an ansatz\footnote{The knowledge of the currents of the classical Boussinesq hierarchy dictates the composite operators appearing in the ansatz. In section 2 of \cite{Ashok:2024zmw}, we have given a list of these currents.} for the current: 
\begin{multline}
    J_{9}(u)= \alpha_{1}(T(T(TW)))(u)+\alpha_{2}(W(WW))(u)+\alpha_{3} (T''W'')(u)\\
    +\alpha_{4}(T'(T'W))(u)+\alpha_{5}(T(WT''))(u)~.
\end{multline}
The stress tensor $T(u)$ has weight 2, the $W(u)$ field has weight 3, while each derivative adds 1 to the conformal weight. One can check that each term in the ansatz for $J_9$ has weight 9. The parentheses indicate the conformal normal ordering of the operators (see Appendix \ref{thermalcorrelators} for details). We note that compared to our analysis in \cite{Ashok:2024zmw}, some of the composite operators have been normal-ordered differently here, primarily for computational convenience. 

\paragraph{2.} The next step involves computing the thermal one-point function of each of the composite operators appearing in the ansatz in a higher spin module:
\be
\langle {\cal O}\rangle = \Tr \, \left(q^{L_0-\frac{c}{24}}\, {\cal O}\right)~,
\ee
where $q = e^{-\beta}$, with $\beta$ being the inverse temperature. The trace is taken in the higher spin module constructed over a highest weight state labeled by the eigenvalues of the commuting operators $(L_0, W_0)$. A generic state in the module will be of the form
\be
\mathcal{P}=\prod_{n>0}L_{-n}^{a_{n}}W_{-n}^{b_{n}}|\Delta_2, \Delta_3\rangle~,
\ee
where we shall associate to every state in the module a level $L$, defined to be 
\be 
L = \sum_{n} n(a_n+b_n)~.
\ee 
We first calculate the thermal correlator of the fields appearing in the composite operator using the Zhu recursion relations, and then normal order them, as described in Appendix \ref{thermalcorrelators}. The thermal one-point functions of each of the composite operators appearing in $J_9$ have been listed in Appendix \ref{listofthermalvevsI8}. It is evident from the expressions that it is $u$-independent. Thus, computing the one-point function of the current $J_9(u)$ is equivalent to computing the one-point function of the conserved charge ${\bf I}_8$. 

\paragraph{3.} At the next step, we take the low-temperature limit of the thermal one-point functions, which essentially amounts to the $q\rightarrow 0$ limit. From the definition of the trace, it is clear that one is effectively computing the eigenvalue of the conserved charge ${\bf I}_8$ in the highest weight state, at level $L=0$. Now, this eigenvalue has already been computed in our previous work \cite{Ashok:2024zmw}, using the ODE/IM correspondence. The result is:
\begin{align}
    I_8&=\langle \Delta_2, \Delta_3| {\bf I}_8| \Delta_2, \Delta_3\rangle \cr
    &=\Delta _2^3\Delta _3 +3\Delta _3^3- \frac{(c+14)}{8}\Delta _2^2\Delta _3 +\frac{(c (5 c+148)+900)}{960}  \Delta _2\Delta _3
\nonumber\\
&\hspace{6cm}-\frac{(c+30) (c (35 c+604)+2940)}{483840}\Delta _3 ~.
\end{align}
By equating the low-temperature limit of the thermal one-point function to this expression, we can fix the current up to one undetermined constant $\gamma$:
\begin{multline}
\label{J9infull}
    J_9=(T(T(TW)))+3(W(WW))\\
   +\frac{1}{32} (7 c+2)(2\pi)^2 (T'(T'W)) -\frac{1}{480} (c + 2) (2 c+55 ) (2\pi)^4 (T''W'')\\
    + (2\pi)^2 \, \gamma\,\left((T(WT''))+\frac{5}{2}(T'(T'W))-\frac{1}{24}(c+25)(2\pi)^2 (T''W'') \right)~.
\end{multline}
A similar analysis can be carried out for the next non-trivial current, which is of weight eleven, and the eigenvalue in the highest weight state fixes the current up to four undetermined constants:
\begin{align}
\label{J11full}
    &J_{11} =(T(T(T(TW))))+6(W(W(WT)))+(2\pi)^2\frac{3(11c+10)}{16}(W^{\prime}(W^{\prime}W))\cr
    &+ (2\pi)^4\frac{(19 c^2+304 c+1492) }{1536}(T^{\prime\prime\prime}( WT^{\prime}))+(2\pi)^6\frac{(480 c^3+12733 c^2+52424 c-93444)}{1935360}(T^{\prime\prime\prime}W^{\prime\prime\prime})\cr
    &+(2\pi)^4\delta_1 \left[(T^{\prime\prime\prime} (WT^{\prime}))+(T^{\prime\prime}(WT^{\prime\prime}))\right]\cr
    &+(2\pi)^4\delta_2 \left((T(WT^{\prime\prime\prime\prime})) +\frac{5}{2}(T^{\prime\prime\prime}( WT^{\prime}))+(2\pi)^2\frac{1}{120} (5 c-21) (T^{\prime\prime\prime}W^{\prime\prime\prime})\right)\cr &+(2\pi)^2\delta_3 \Big( (T(T(T^{\prime\prime}W)))+\frac{21}{2}(W^{\prime}(W^{\prime}W))
    +\frac{(2\pi)^2}{192} (19 c-318)(T^{\prime\prime\prime}( WT^{\prime})) \cr
    &\hspace{9cm} +
    (2\pi)^4\frac{20c^2-11c+2318}{11520}(T^{\prime\prime\prime}W^{\prime\prime\prime})\Big) \cr
    &+(2\pi)^2\delta_4 \Big(T^{\prime}(T^{\prime}(TW)))
    -3(W^{\prime}(W^{\prime}W)) -\frac{1}{96}(2\pi)^2(c-198)(T^{\prime\prime\prime}( WT^{\prime})) \cr
    &\hspace{10cm}+ 
    (2\pi)^4\frac{101c-598}{5760}(T^{\prime\prime\prime}W^{\prime\prime\prime})\Big)  \,.  
\end{align}
As stressed in our previous work, the combination of composite operators multiplying the coefficients $\gamma$ and $\delta_i$ in the currents is actually trace-free. This means that their one-point functions are identically zero (and not just the leading term as $q\rightarrow 0$).  
We also note that due to the reordering in some of the composite operators, the trace-free combinations have coefficients that differ from those in \cite{Ashok:2024zmw}. 

\paragraph{4.} The next step is to fix the leading dependence of the coefficients $\gamma$ and $\delta_i$ on the central charge of the theory. For this purpose we take the classical limit, which amounts to taking the large-$c$ limit. The precise way to do this has been outlined in  \cite{Bazhanov:2001xm}; we re-scale and redefine
\be 
T(u)\longrightarrow \left(\frac{c}{24}\right) U(u)~,\qquad W(u) \longrightarrow i \left(-\frac{c}{24}\right)^{\frac32} V(u) ~,
\ee
while simultaneously re-scaling the quantum Boussinesq charges in the following manner:
\begin{align} 
{\bf I}_k \longrightarrow i^{-k-1}\left(-\frac{c}{24}\right)^{\frac{k+1}{2}}\, I_{k}^{\text{class}}~, 
\end{align}
in the limit that $c\rightarrow -\infty$. The classical charges on the right-hand side are the conserved charges of the classical Boussinesq hierarchy, as listed in \cite{Ashok:2024zmw}. Specifically, for $k=8$ and $k=10$, these charges are given by 
\begin{align}
I_{8}^{\text{class}} =&\int dx~ \big( U^3 V +3 V^3-3 U V U''-\frac{9}{4} V U'^2+\frac{3}{5}U'' V''\big)~,\\
 I_{10}^{\text{class}}=&\int dx~\big(U^4 V+6 U V^3+9 V V'^{2}-6 U^{2} V U''-\frac{15}{2}U V U'^{2}+\frac{21}{4}V U''^{2}\notag\\&\hspace{3.8cm}+\frac{15}{2}V U' U^{(3)}+3 U V U^{(4)}+\frac{3}{7}V^{(3)}U^{(3)}\big) ~.
\label{I8classical}~
\end{align}
These relations constrain the form of the undetermined constants to 
\be
\label{cexpofcoefficients}
\begin{aligned} 
\gamma &= -\frac{c}{8} + \gamma^{(0)}\\
\delta_1 &=\phantom{-}\frac{7c^2}{768} + \delta_1^{(1)}c + \delta_1^{(0)}~,\qquad 
\delta_2 =\frac{c^2}{192} + \delta_2^{(1)}c + \delta_2^{(0)} ~,\\
\delta_3 &=-\frac{c}{4} + \delta_3^{(0)} ~,\hspace{2.5cm}
\delta_4 = -\frac{5c}{16} + \delta_4^{(0)}~.
\end{aligned}
\ee
One of the main goals of this paper is to determine the values of the remaining unknown coefficients in \eqref{cexpofcoefficients}.

\paragraph{5.} To compute the undetermined constants, it is clear that we have to go beyond the one-point functions we have calculated so far. As shown in \cite{Ashok:2024zmw}, a natural idea is to consider the two-point function involving the conserved charge and another operator, which we denote by ${\bf A}$: 
\be 
\label{2ptfunction}
 \langle {\bf I}_n\, {\bf A}\rangle =\Tr\left(q^{L_0-\frac{c}{24}}\,{\bf I}_n\, {\bf A}  \right) ~.
\ee
We shall choose ${\bf A}$ to be the zero mode of a composite operator built out of the stress tensor and/or the spin-3 current. For the undetermined coefficients to appear in the correlator \eqref{2ptfunction}, the operator ${\bf A}$ must be chosen so that its two-point correlator with the trace-free combinations appearing in the currents is non-vanishing\footnote{Such an operator is likely to exist due to the trace being non-degenerate (when considered as a bi-linear form), and the fact that there are infinitely many linearly independent local operators to choose from.}. The two-point function can be calculated by taking, in turn, each composite operator in $J_9$ and $J_{11}$ and computing the two-point function with ${\bf A}$ by using the Zhu recursion relations. This is a modification of the calculations we performed for the one-point functions, and the results for the $J_9$ composites are presented in Appendix \ref{J9twopoint}. 

\paragraph{6.} We now compute the two-point function in a completely independent manner by using the operator formalism. To facilitate this, we make a simple choice for ${\bf A}$ so that its action on states in the higher spin module is well understood. However, it is crucial that ${\bf A}$ be the zero mode of a composite operator. Being the zero mode ensures that ${\bf A}$ acts as a linear operator on the level-$L$ excited states of the higher spin module, just like the conserved charge itself. Thus, one can restrict attention to states at a given level and consider the contribution to the trace \eqref{2ptfunction} from the level-$L$ excited states. Let $|\psi_i\rangle$ denote the usual basis elements of weight $\Delta_2+L$ in the higher spin module, built out of the $L_{-m}$ and $W_{-n}$ operators. Then, one can straightforwardly compute 
\be 
{\bf A}\, |\psi_i \rangle = \sum_{j}\, A_{ji}\,  |\psi_j\rangle~,
\ee 
by using the ${\cal W}_3$ algebra. Now, at each level, the mutually commuting conserved charges ${\bf I}_n$ can be diagonalized, and we have corresponding eigenstates $|e_{i}\rangle$ such that
\be 
{\bf I}_n\, | e_i\rangle =  I_{n,i}^{(L)}\, | e_i\rangle~.
\ee 
The basis elements $|\psi_i\rangle$ are obviously linear combinations of these eigenstates:

\be
\label{Rdefn}
|\psi_j\rangle = \sum_{k} \, R_{kj}|e_k\rangle ~.
\ee
It is now a matter of simple algebra to check that the contribution to the trace in \eqref{2ptfunction} at level $L$ is  given by
 \be
 \label{levelLtrace}
\langle {\bf I}_n\, {\bf A}\rangle_{{\cal H}_{L}}  =  q^{\Delta_2+L}\, \Tr\left( A\, R\, I_{n}^{(L)}\, (R^{-1})\right)~.
\ee
The matrices $A$ and $R$ can be obtained using the ${\cal W}_3$ algebra, as we shall show in the following section. Thus, the one remaining ingredient that cannot be directly calculated in the operator formalism is the set of eigenvalues of the Boussinesq charges at a given level. 

\paragraph{7.} The eigenvalues at a given level in the higher spin module are obtained via the ODE/IM correspondence \cite{Dorey:1998pt, Bazhanov:1998wj, Masoero:2018rel, Masoero:2019wqf}. Since this has been discussed in detail in our previous work, we shall be brief in our analysis in Section \ref{level2eigenvalues}.

\paragraph{8.} With this final piece of the puzzle in place, we compare and match the result in \eqref{levelLtrace} with the $O(q^{\Delta_2+L})$ contribution to the trace computed using the Zhu recursion, as explained in point 5 above. By equating the two results, the idea is to unambiguously fix the undetermined coefficients in \eqref{cexpofcoefficients}. A crucial point here is that the operator ${\bf A}$ must necessarily have a non-vanishing two-point function with the trace-free combinations appearing in the currents.  This ensures that equating the trace computed in two different ways leads to simple linear equations for the coefficients appearing in the ansatz.

\section{The Operator Formalism}
\label{operatorformalism}
In this section, we extend the analysis to the second excited level and outline the derivation of the matrices $A$ and $R$ required to compute the trace in \eqref{levelLtrace}. Since we make use of the ${\cal W}_3$ algebra for this purpose, we provide the commutation relations between the modes $L_n$ of the spin-2 current and the modes $W_n$ of the spin-3 current,  as presented in   \cite{Zamolodchikov:1985wn}: 
\begin{align*}
[L_n,L_m]  =& (n-m)L_{n+m} + \frac{c}{12}(n^3-n)\delta_{m+n, 0}~,\qquad  [L_n, W_m] = (2n-m)W_{n+m} ~, \\
[W_n,W_m] =& \frac{(n-m)}{3}\Lambda_{n+m}+\frac{n-m}{3b^2}\left(\frac{1}{15}(n+m+3)(n+m+2) -\frac16(n+2)(m+2) \right)L_{n+m}\nonumber \\
&\hspace{2.8cm}+\frac{c}{1080b^2}n(n^2-4)(n^2-1)\delta_{n+m,0}~. \end{align*}
Here $c$ is the central charge and $b^2 = \frac{16}{5c+22}$. The composite operator appearing on the right-hand side of the $W_n$ commutators is the normal ordered operator $\Lambda(u)= (TT)(u)$, whose modes are given by  
\be 
\Lambda_n = \sum_{k=-\infty}^{\infty} :L_kL_{n-k}: + \frac{1}{5}x_n L_n~,
\ee
where $x_{2l} = 1-l^2$ and $x_{2l+1} = 2-l-l^2$. The normal ordering symbol $:\,:$ indicates that we put the operators with larger index $n$ to the right. 

\subsection{Level 1 matrices}

Most relevant calculations at level one have already been done in \cite{Ashok:2024zmw}, so we will simply present the results and refer the reader to that reference for the details.
The basis vectors at level 1 are given by
\be 
|\psi\rangle \in \left\{ L_{-1}|\Delta_2, \Delta_3\rangle, W_{-1}|\Delta_2, \Delta_3\rangle \right\}~.
\ee 
The eigenstates of the mutually commuting charges are written in terms of the basis vectors as follows: 
\begin{equation}
\label{leveloneEVs}
\begin{aligned}
    \rvert e_1 \rangle &=  -\frac{\sqrt{-c+32 \Delta _2+2}}{4 \sqrt{6}} L_{-1}\rvert \Delta_2,\Delta_3 \rangle + W_{-1} \rvert \Delta_2,\Delta_3 \rangle~, \\  \rvert e_2 \rangle &=  \frac{\sqrt{-c+32 \Delta _2+2}}{4 \sqrt{6}} L_{-1}\rvert \Delta_2,\Delta_3 \rangle + W_{-1} \rvert \Delta_2,\Delta_3 \rangle~.
\end{aligned}
\end{equation}
As shown in \cite{Ashok:2024zmw}, this is derived from the action of $W_0$ on the level-1 basis, and by diagonalization. The inverse relation fixes the form of the R-matrix defined in \eqref{Rdefn}:
\begin{align}
    R = \left(
\begin{array}{cc}
 -\frac{2 \sqrt{6}}{\sqrt{-c+32 \Delta _2+2}} &  \frac{1}{2} \\
 \frac{2 \sqrt{6}}{\sqrt{-c+32 \Delta _2+2}} & \frac{1}{2} \\
\end{array}
\right)~.
\end{align}

In our attempt to fix the form of the currents $J_9$ and $J_{11}$, we will find it useful to consider two different ${\bf A}$-operators:
\begin{align}
 {\bf A}_3 &=\int_0^1\, du\, (TT)(u) = 2\sum_{n\ge0}L_{-n}L_{n}+ L_{0}^{2}-\frac{c+2}{12}L_{0}+\frac{c(5c+22)}{2880}~,\\
{\bf A}_5 &= (2\pi)^2\,\int_0^1\, du\,  (T'T')(u)=  2\sum_{n>0} n^2  L_{-n} L_{n} + \frac{1}{60} L_0 -\frac{31c}{30240}~.
\end{align}
The subscript indicates the weight of each operator. 
At level 1, it is straightforward to find their action on the basis states using the commutation relations of the ${\cal W}_3$ algebra. The results for ${\bf A}_3$ have already been carried out in our previous work. We present the results for ${\bf A}_5$ here\footnote{For the computation of the matrix elements, we have used an implementation of $\mathcal{W}_3$ algebra in Mathematica. For this purpose, we have used the Mathematica notebook \href {https://sites.google.com/view/matthew-headrick/mathematica}{Virasoro} by Matthew Headrick with minor modifications.}
\begin{align}
(A_5)_{ij} &= 
        \left(
\begin{array}{cc}
 -\frac{31 c}{30240}+\frac{241 \Delta _2}{60}+\frac{1}{60} & 6 \Delta _3  \\
 0& \frac{1}{60} \left(\Delta _2+1\right)-\frac{31 c}{30240} \\
\end{array}
\right)\,.
\end{align}
Substituting these matrices, we find that the level-1 contribution to the traces from the operator formalism can be written compactly:
\begin{align}
       \langle {\bf I}_n\, {\bf A}_5\rangle_{{\cal H}_{1}}  &=  q^{\Delta_2+1}\,
       \left\{ \left(\frac{-31 c+60984 \Delta _2+504}{30240}\right)(I_{n,1}^{(1)}+I_{n,2}^{(1)})\right. \cr&\left.\hspace{7cm}+\frac{12 \sqrt{6} \Delta _3}{\sqrt{-c+32 \Delta _2+2}}(I_{n,2}^{(1)} -I_{n,1}^{(1)} ) \right\}~.
\end{align}
The result for the trace with ${\bf A}_3$ inserted is\cite{Ashok:2024zmw}:
\begin{align}
       \langle {\bf I}_n\, {\bf A}_3\rangle_{{\cal H}_{1}}  &=  q^{\Delta_2+1}\,\left\{ \left(\Delta _2^2 +\frac{\left(5 c^2-218 c+2400\right) }{2880}-\frac{1}{12} (c-46) \Delta _2\right)( I_{n,1}^{(1)}+I_{n,2}^{(1)}) \right. \nonumber\\  &\left.\hspace{7cm}+\frac{12 \sqrt{6} \Delta _3 }{\sqrt{-c+32 \Delta _2+2}}(I_{n,2}^{(1)} -I_{n,1}^{(1)} ) \right\}~.
\end{align}
What is left to compute are the eigenvalues $I_{n,i}^{(1)}$, which can be computed using the ODE/IM correspondence, as shown in detail in \cite{Ashok:2024zmw}. With this in place, the trace with the ${\bf A}$-operators insertion can be calculated explicitly from the operator formalism for any of the Boussinesq charges.

\subsection{Level 2 matrices}

The basis vectors at the $L=2$ level are given by
\be 
\label{level2basis}
\psi \in \left\{L_{-2}|\Delta_2, \Delta_3\rangle, L_{-1}^2|\Delta_2, \Delta_3\rangle, L_{-1}W_{-1}|\Delta_2, \Delta_3\rangle, W_{-1}^2|\Delta_2, \Delta_3\rangle, W_{-2}|\Delta_2, \Delta_3\rangle \right\}~.
\ee
To find the eigenstates of the mutually commuting charges, we begin with the $5\times 5$ matrix representation of the simplest non-trivial conserved charge ${\bf I}_2 = W_0$, given by 
\begin{equation}
\label{W0inlevel2basis}
    \left(
\begin{array}{ccccc}
 \Delta _3 & 0 & \frac{1}{15} \left(-\frac{1}{b^2}+10 \Delta _2+12\right) & 0 & \frac{2}{3} \\
 0 & \Delta _3 & \frac{4 \Delta _2}{3} & 2 \Delta _3 & \frac{4 \Delta _2}{3} \\
 4 & 0 & \Delta _3 & \frac{2}{15} \left(-\frac{1}{b^2}+10 \Delta _2+7\right) & 0 \\
 0 & 0 & 2 & \Delta _3 & 0 \\
 -2 & 4 & 0 & \frac{1}{15} \left(\frac{1}{b^2}-10 \Delta _2-12\right) & \Delta _3 \\
\end{array}
\right)\,.
\end{equation}
This is easily derived by acting with $W_0$ on the basis of vectors using the basic commutation relations.  
The eigenvectors at $L=2$ are obtained as column vectors of the matrix used to diagonalize this matrix. This will also automatically provide the form of the R-matrix, as the similarity transformation that diagonalizes this matrix. Since we have a general quintic equation to solve, unlike the $L=1$ case, it is not possible to obtain a closed-form expression for the eigenvectors at the second excited level. However, for a given numerical choice of $(c, \Delta_2, \Delta_3)$ this is easily done, and this is all we shall need to fix the form of the currents, as we shall explain in the subsequent analysis. 

In the basis \eqref{level2basis}, one can similarly compute the matrix representation of ${\bf A}_5$:

{\scriptsize
\begin{align}
    \left(
\begin{array}{ccccc}
 -\frac{31 c}{30240}+\frac{481 \Delta _2}{60}+\frac{121}{30} & 6 & 6 \Delta _3 & -\frac{2}{5 b^2}+4 \Delta _2+\frac{4}{5} & 0 \\
 48 \Delta _2 & \frac{120929 c}{30240}+\frac{1921 \Delta _2}{60}+\frac{1}{30} & 120 \Delta _3 & \frac{8 \Delta _2 \left(10 b^2 \Delta _2+2 b^2-1\right)}{3 b^2} & 48 \Delta _3 \\
 0 & 0 & -\frac{31 c}{30240}+\frac{241 \Delta _2}{60}+\frac{361}{30} & 12 \Delta _3 & 8 \\
 0 & 0 & 0 & -\frac{31 c}{30240}+\frac{\Delta _2}{60}+\frac{1}{30} & 0 \\
 0 & 0 & -4 \left(\Delta _2+3\right) & -12 \Delta _3 & -\frac{31 c}{30240}+\frac{\Delta _2}{60}-\frac{239}{30} \\
\end{array}
\right)
\end{align}
}
and a similar matrix can be computed for ${\bf A}_3$ as well. Thus, all that remains is to calculate the eigenvalues of the Boussinesq charges at the second excited level. 

\section{ODEs for Excited State Eigenvalues}
\label{level2eigenvalues}

We now review the results of \cite{Masoero:2018rel, Masoero:2019wqf}, in which a class of third-order ordinary differential equations have been proposed, and that can be used to compute the eigenvalues of the quantum Boussinesq charges in higher spin modules. The ODE is given by
\begin{equation}\label{beq:ode1}
	\phi ^{(3)}(z)-W_1(z) \phi '(z)+W_2(z) \phi (z)=0~,
\end{equation}
with $W_{1}$ and $W_{2}$ given by,
\begin{eqnarray}
W_{1}&=&\frac{\bar{r}_1}{z^2}+\sum _{j=1}^L \left(\frac{a^{(j)}_{11}}{z \left(z-w_j\right)}+\frac{3}{\left(z-w_j\right){}^2}\right) ~,\\
W_{2}&=&\frac{\bar{r}_2}{z^3}+\frac{1}{z^2}+\lambda 
z^k+\sum _{j=1}^L \left(\frac{a^{(j)}_{22}}{z^2 \left(z-w_j\right)}+\frac{a^{(j)}_{21}}{z \left(z-w_j\right){}^2}+\frac{3}{\left(z-w_j\right){}^3}\right)~.
\end{eqnarray} 
Here $L$ refers to the level of the excited state. Setting $k=-\frac{3M+2}{M+1}$, the parameters $(M, {\bar r}_1, {\bar r}_2)$ are related to the central charge $c$ and the dimensions $(\Delta_2, \Delta_3)$ of the conformal field theory 
\be 
\label{ODEtoCFTmap}
\begin{aligned}
             c&= 2-\frac{24M^{2}}{(M+1)}~,\\
    \bar{r}_{1}&= \frac{9 M^2+9 \Delta _2 (M+1)-9 (M+1)^2}{9 (M+1)^2}~,\\
    \bar{r}_{2}&= \frac{M^2}{(M+1)^2}+\frac{\Delta _2}{M+1}-\frac{\Delta _3}{(M+1)^{3/2}}-1~.
\end{aligned}
\ee
There are $4L$ further parameters $(a_{11}^{(j)}, a_{21}^{(j)}, a_{22}^{(j)}, w_j)$, with $j=1, \ldots L$, that appear in the ODE. These are determined by imposing that there is trivial monodromy around each $w_j$ \cite{Bazhanov:2003ni, Masoero:2018rel, Masoero:2019wqf}. This leads to algebraic constraints between the $4L$ parameters. $2L$ of these are immediately determined in terms of the rest as follows:
\begin{align}\label{monoconst1}
 a^{(j)}_{11}&= k~,\qquad 
 a^{(j)}_{22}= -\frac{k^{2}}{3} + a^{(j)}_{21} +\frac{2}{3} k a^{(j)}_{21}~.
\end{align}
Renaming $ a^{(j)}_{21}=a_{j}$, the remaining $2L$ parameters $(a_j, w_j)$ satisfy the non-trivial constraints \cite{Masoero:2019wqf}:
\begin{align}
    a_\ell^2-ka_\ell+k^2+3k-3\bar{r}^1&=\sum_{\substack{j=1\\ j\neq \ell}}^L\left(\frac{9w_\ell^2}{(w_\ell-w_j)^2}+\frac{3kw_\ell}{w_\ell-w_j}\right),\cr
\mathcal{A}a_\ell+\mathcal{B}-9(k+2)w_\ell&=\sum_{\substack{j=1\\j\neq \ell}}^L\left(\frac{18(-k+a_\ell+a_j)w_\ell^3}
{(w_\ell-w_j)^3}+\frac{(3 w_\ell^2 \left(k \left(3 a_\ell-4 k-3\right)+a_j (5 k+6)\right)}{(w_\ell-w_j)^2}\right. \cr
&\hspace{2cm}-\left.\frac{(2 k+3) w_\ell \left(k \left(-a_\ell+2 k+3\right)-3 a_j (k+2)\right)}{w_\ell-w_j}\right).\label{monoconst2}
\end{align}
Here the $\mathcal{A}$ and $\mathcal{B}$ are given in terms of the CFT data by
  \begin{align}
  \mathcal{A}= &(18+2 k^3+14 k^2-2 k \bar{r}_1+30 k-6\bar{r}_1)~,\\
     \mathcal{B}=& (k+3) \left((k+9) \bar{r}_1-k (k+1) (k+3)-9 \bar{r}_2\right)~.
     \end{align}
The non-trivial result of \cite{Masoero:2018rel, Masoero:2019wqf} is that there are exactly as many solutions to these algebraic constraints as the number of states at the $L$-th excited level in the higher spin module. 

For the first excited level, the procedure to solve the ODE \eqref{beq:ode1} and extract the eigenvalues of the conserved charges has already been discussed in detail in \cite{Ashok:2024zmw}. In this section we only highlight the distinct features of the ODE \eqref{beq:ode1} for the second excited states. The main change going from level one ODE to level two ODE is in the nature of the constraint equations for $\{a_{1},w_{1},a_{2},w_{2}\}$ \eqref{monoconst2}. As we can see from the equations, up to level one, the contribution from the right-hand side vanishes. However, for levels two and beyond, we cannot solve these equations analytically due to the non-trivial contributions from the right-hand side. Still, we can proceed numerically and solve these equations for particular choices of ODE parameters $\{k,\bar{r}_{1},\bar{r}_{2}\}$ (or, equivalently, for a given central charge and conformal dimensions $(c, \Delta_2, \Delta_3)$).

Other than this non-triviality with the constraint equations, the WKB analysis for the level two case remains the same as the level 1 case, and the level $L$ eigenvalues can be found from the WKB periods:
\begin{equation}
    \widehat{{I}}^{(L)}_{n} 
    = \oint dz\, S^{(L)}_{n+1}(z) \,. 
\end{equation}
For $L=2$, the $S^{(2)}_{n}$'s are related to the functions that appear at the $n$-th order in the WKB ansatz for the wave function that solves the third order ODE. There are multiple ways to calculate the period integrals (see the recent paper \cite{Ito:2024kza} for a recursive way to perform the period integrals in the vacuum case). We calculate these period integrals following the methods outlined for the vacuum and first excited level in \cite{Ashok:2024zmw}, and these are related to the eigenvalues of the conserved charge ${\bf I}_n$ in one of the $L$-th excited states:
\begin{equation}\label{eq:EV2}
    \widehat{{I}}^{(L)}_n =c^{(L)}_n~ e^{-\frac{n\pi i}{3} } (1+M)^{\frac{n+1}{2}}\frac{\Gamma(\frac{n}{3M} + \frac{n}{3})\Gamma(-\frac{n}{3}) }{\Gamma(\frac{n}{3M})}~{I}^{(L)}_n~. 
\end{equation}
A crucial point here is that the relative numerical factor between the period integrals and the eigenvalues turns out to be exactly the same\footnote{Getting this numerical prefactor correct is crucial especially in light of the work done in Appendix \ref{fixingJ9}, where the eigenvalues are computed numerically for specific choices of $(c, \Delta_2,\Delta_3)$.} as it was for the vacuum case (up to a sign $c_{n}^{(2)}=(-1)^{n}c_{n}^{(0)}$).  

We list the expressions of the first few excited state eigenvalues at the second level:
\begin{align}
    I_1^{(2)} &= \Delta_2 -\frac{c}{24} +2 \label{I1level2}\,,\\
    I_2^{(2)} &= \Delta_3-\frac{2 (3 M+2)}{3 \sqrt{M+1}}-\frac{2}{3} \sqrt{M+1}(a_1+a_2)\,,\label{I2level2}\\
    I_4^{(2)} &= \frac{\Delta _3 \left(3 M^2+5 M+5\right)}{3 (M+1)}+\Delta _2 \left(\Delta _3-\frac{2 (3 M+2)}{3 \sqrt{M+1}}\right)-\frac{2 (3 M+2) \left(3 M^2+5 M+5\right)}{9 (M+1)^{3/2}}\cr
    &~+\left(-\frac{2 \left(3 M^2+5 M+5\right)}{9 \sqrt{M+1}}-\frac{2}{3} \Delta _2 \sqrt{M+1}\right)(a_1+a_2)-3 M (M+1)^{3/2}(w_1+w_2)~.\label{I4level2}
\end{align}
The $(a_i, w_i)$ that appear in the above expressions are solutions to the algebraic constraint equations. Although the eigenvalues of the higher charges can be similarly calculated, the expressions get increasingly complicated for higher charges, so we do not list them. We will be concerned with the eigenvalues $I_{8}^{(2)}$ and $I_{10}^{(2)}$ respectively.  As shown in Appendix \ref{fixingJ9}, we will compute these numerically for particular values of $(c, \Delta_2, \Delta_3)$. 

\section{The Higher Currents}
\label{highercurrents}

We now have all the ingredients in place to proceed with the determination of the higher-weight currents.  We will use a combination of analytical and numerical methods to fix the undetermined coefficients. The thermal correlators are computed analytically, while the eigenvalues, as well as the $R$ matrices in the operator formalism, are computed numerically for particular values of $(c, \Delta_2, \Delta_3)$.

Let us first discuss the case of the current $J_9$. We recall from \eqref{J9infull} and \eqref{cexpofcoefficients} that the form of the current is given by: 
\begin{multline}
    J_9(u) = (T(T(TW)))(u)-\frac34(2\pi)^2(T(WT''))(u) +\frac{1}{32} (7 c-58)(2\pi)^2 (T'(T'W))(u)\\
   +3(W(WW))(u)-\frac{1}{480} (2c^2+44c-265) (2\pi)^4 (T''W'')(u) + \left(-\frac{c}{8} + \gamma^{(0)}\right) {\cal F}(u)~,
\end{multline}
where we have denoted the trace-free combination appearing in \eqref{J9infull} by ${\cal F}(u)$.  The goal is to determine one $c$-independent coefficient $\gamma^{(0)}$ to completely fix the current.

We first compute the trace with the insertion of ${\bf A}_3$. However, we find that the two-point function
\begin{align}
     \Big\langle {\bf A}_3\, {\cal F}(u) \Big\rangle = 0~.
\end{align}
Thus, the two-point function of ${\bf A}_3$ with ${\bf I}_8$ cannot be used to determine the coefficient $\gamma^{(0)}$. We next try with the insertion of ${\bf A}_5$; the results are given in Appendix \ref{J9twopoint}. We find that
\begin{equation}
    \Big\langle 
    {\bf A}_5\, {\cal F}(u)
    \Big\rangle
   = 216(c+2)\Delta_3\, q^{\Delta_2-\frac{c}{24}}\, ( q^2 + 58q^3  + \ldots)~ .
\end{equation}
We see that there is a non-trivial contribution at $O(q^2)$. Thus, if we can compute the contribution to the trace in \eqref{levelLtrace} from the second excited level $L=2$ in the operator formalism, it would be possible to match with the result for the trace from the thermal two-point function and determine $\gamma^{(0)}$. 

Since the expressions are rather cumbersome, we have relegated the details of the calculation to the appendices. The thermal two-point functions of ${\bf A}_5$ with each of the composite operators in the ansatz for $J_9$ are given in Appendix \ref{J9twopoint}, while the steps leading to the calculation of the eigenvalues of ${\bf I}_8$ at the second excited level have already been given in Section \ref{level2eigenvalues}. The matching of the two sets of results is given in Appendix \ref{fixingJ9}. From the results summarized in Table \ref{table:1}, we deduce that the coefficient $\gamma^{(0)}$ takes the simple value 
\be 
\gamma^{(0)} = -\frac{3}{4}~.
\ee
This fixes the form of the current density to be
\begin{multline}
    J_9 = (T(T(TW)))+3(W(WW))-\frac34(2\pi)^2(T(WT''))\\
   +\frac{1}{32} (7 c-58)(2\pi)^2 (T'(T'W)) -\frac{1}{480} (2c^2+44c-265) (2\pi)^4 (T''W'')~.
\end{multline}

We next turn to fix the undetermined coefficients appearing in the weight-11 current. We begin by recalling the form of the current that we obtained after using the low-temperature limit of the one-point function and the classical limit: 
\begin{align}
    &J_{11} =(T(T(T(TW))))+6(W(W(WT)))+(2\pi)^2\frac{3(11c+10)}{16}(W(W^{\prime}W^{\prime}))\cr
    &+ (2\pi)^4\frac{(19 c^2+832 c+1972) }{1536}(T^{\prime\prime\prime}( WT^{\prime}))+(2\pi)^6\frac{(120 c^3+6937 c^2+24719 c-15906)}{483840}(T^{\prime\prime\prime}W^{\prime\prime\prime})\cr
    &+\left(\frac{7c^2}{768} + \delta_1^{(1)}c + \delta_1^{(0)}\right)\, {\cal G}_1 + 
\left(\frac{c^2}{192} + \delta_2^{(1)}c + \delta_2^{(0)}\right) \, {\cal G}_2 + 
\left(-\frac{c}{4} + \delta_3^{(0)}\right)\, {\cal G}_3 + \left(-\frac{5c}{16} + \delta_4^{(0)}\right)\, {\cal G}_4~.\cr
\label{J11afterclassical}
\end{align}
We have denoted by ${\cal G}_i$ the four trace-free combinations appearing in \eqref{J11full}. 

We first compute the two-point functions of the ${\cal G}_i$ with the operators ${\bf A}_j$ to identify the level at which the eigenvalues of ${\bf I}_{10}$ must be calculated to determine the coefficients.

The complete expressions for thermal correlators are rather cumbersome, and we present here only the leading terms in the $q$-expansion. We first list the two-point function of the ${\cal G}_i$ with ${\bf A}_5$: 
\begin{align*}
    \Big\langle {\bf A}_5\, {\cal G}_i \Big\rangle =\begin{cases}
      q^{\Delta_2-\frac{c}{24}+2}\Big(-96 (2 c+1) \Delta _3-288 \Delta _2 \Delta _3  \Big) + \ldots & i=1 \\
      \\
       O(q^3) & i=2 \\
       \\
     20160 \Delta _3 q^{\Delta_2-\frac{c}{24}+1} \Big((c-2) \Delta _2^2-32 \Delta _2^3+216 \Delta _3^2\Big)  &i=3\\
    -5760 q^{\Delta_2-\frac{c}{24}+2}\Big(-51408 \Delta _3^3+7616 \Delta _3 \Delta _2^3+(1778 c+5516) \Delta _3 \Delta _2^2\\
    \hspace{.5cm}+\left(112 c^2+4805 c+4246\right) \Delta _3 \Delta _2+\left(60 c^2+1715 c+6202\right) \Delta _3 \Big)  +\ldots    \\
    \\
     -2\Delta_3q^{\Delta_2-\frac{c}{24}+1}\Big(\left(c-2\right)  \Delta _2^2-32\Delta _2^3+216 \Delta _3^2\Big) & i=4 \\
     \hspace{.5cm} +q^{\Delta_2-\frac{c}{24}+2}\Big(1088 \Delta _2^3 \Delta _3-7344 \Delta _3^3+(254 c+788) \Delta _2^2 \Delta _3
      \\ 
      \hspace{1.5cm}+\left(16 c^2+779 c+1522\right) \Delta _2 \Delta _3+\left(6 c^2+809 c+1294\right) \Delta _3\Big) + \ldots
    \end{cases}
\end{align*}
We make a few observations about these results. Firstly, the two-point function of ${\bf A}_5$ with ${\cal G}_2$ is zero up to the second level. So, this thermal two-point function will not be able to fix $\delta_2^{(0)}$ and $\delta_2^{(1)}$. Secondly, the $O(q)$ contribution to the two-point function of ${\bf A}_5$ with ${\cal G}_3$ and ${\cal G}_4$ are equal up to a numerical coefficient. So, this should give a linear relation between $\delta_3^{(0)}$ and $\delta_4^{(0)}$. However, by setting different values of $(c, \Delta_2, \Delta_3)$ in the $O(q^2)$ contribution to the trace, one should be able to find linearly independent equations to fix four of the six undetermined coefficients by comparing with the results of the trace from the operator formalism. 
The specific numerical details of their derivation are provided in Appendix \ref{fixingJ9}.

Since we need more constraints to fix the constants, we move on to consider the two-point function of the ${\cal G}_i$ with ${\bf A}_3$. This is given by
\begin{align*}
    \Big\langle {\bf A}_3\, {\cal G}_i \Big\rangle =\begin{cases}
      q^{\Delta_2-\frac{c}{24}+2}\Big(-48 (c+2) \Delta _3  \Big) + \ldots & i=1 \\
      \\
       q^{\Delta_2-\frac{c}{24}+2}\Big(216 (c+2) \Delta _3 \Big) + \ldots& i=2 \\
       \\
     20160 \Delta _3 q^{\Delta_2-\frac{c}{24}+1} \Big((c-2) \Delta _2^2-32 \Delta _2^3+216 \Delta _3^2\Big) & i=3\\
     -2880 \times 14   q^{\Delta_2-\frac{c}{24}+2}\Big(  320 \Delta _2^3\Delta _3-2160 \Delta _3^3 +(62 c+308) \Delta _2^2\Delta _3  \\ 
     \hspace{.75cm}+  \left(4 c^2+218 c-92\right) \Delta _2\Delta _3 + \frac{1}{14} (-15 c^2+1540 c+6500 )\Delta _3\Big) + \ldots &\\
     \\
      -2\Delta_3q^{\Delta_2-\frac{c}{24}+1}\Big(\left(c-2\right)  \Delta _2^2-32\Delta _2^3+216 \Delta _3^2\Big)& i=4  \\
     \hspace{.75cm} +q^{\Delta_2-\frac{c}{24}+2}\Big(320 \Delta _2^3 \Delta _3-2160 \Delta _3^3 +(62 c+308) \Delta _2^2 \Delta _3 \\ 
      \hspace{1.25cm}+\left(4 c^2+218 c-92\right) \Delta _2 \Delta _3 +\left(\frac{3 c^2}{2}+404 c+1042\right) \Delta _3\Big) + \ldots
    \end{cases}
\end{align*}
We note that the two-point functions of ${\bf A}_3$ with ${\cal G}_1$ and ${\cal G}_2$ are proportional to each other by a purely numerical factor. In fact, one can check that this is true for all orders in the $q$-expansion. Interestingly, the $O(q)$ contribution to the two-point function of ${\bf A}_3$ with ${\cal G}_3$ and ${\cal G}_4$ are identical to those that appear in the two-point function with ${\cal A}_5$ (including the numerical factors). So these do not give any new constraints on the coefficients. However, the $O(q^2)$ contribution for $i=2$ does appear, and it turns out that the two-point function at level 2 is sufficient to fix the remaining two coefficients.  We find 
\be 
\begin{aligned}
    \delta_1^{(1)} &=~-\frac{253}{192},  \quad \delta_1^{(0)} = -\frac{879}{64}\,,  \\
    \delta_2^{(1)} &=\phantom{-}\frac{11}{96} ~, \quad \delta_2^{(0)} =\phantom{-} \frac{43}{12}\,,\\
    \delta_3^{(0)} &=-\frac{43}{8} ~, \quad \delta_4^{(0)} = -\frac{3}{2}~.
\end{aligned}
\ee
The current $J_{11}$ therefore takes the form,
\begin{align}
\label{J11final}
    &J_{11} =(T(T(T(TW))))+6(W(W(WT)))-\frac{5c+86}{16}(2\pi)^2 (T^{\prime}(T^{\prime}(T W)))\cr
    &-\frac{ (c+6)}{8}(2\pi)^2\Big(2 (T(T(T^{\prime\prime} W)))-3(W(W^{\prime}W^{\prime}))\Big)+\frac{ \left(c^2+22 c+688\right)}{192} (2\pi)^4(T(WT^{\prime\prime\prime\prime}))\cr
    &+\frac{\left(7 c^2-1012 c-10548\right)}{768} (2\pi)^4(T^{\prime\prime}(WT^{\prime\prime}))+\frac{ \left(5 c^2-444 c-4764\right)}{384} (2\pi)^4 (T^{\prime\prime\prime}(WT^{\prime}))\cr
    &\hspace{5cm}+\frac{\left(30 c^3+2513 c^2+44188 c-405612\right)}{967680}(2\pi)^6(T^{\prime\prime\prime}W^{\prime\prime\prime})~.
\end{align}
This completes the determination of the higher currents $J_9$ and $J_{11}$ of the Boussinesq hierarchy.

\section{Summary and Discussion}

The main result of this work is the derivation of the currents  $J_9$ and $J_{11}$ of the quantum Boussinesq hierarchy. This was made possible by bringing together two completely different approaches to the study of the underlying integrable system. On the one hand, we used conformal field theory methods, specifically the Zhu recursion relation, to calculate the thermal correlation functions of the stress tensor and the spin-3 current in a higher spin module. On the other hand, we made use of the ODE/IM correspondence to calculate the eigenvalues of the quantum Boussinesq charges in the excited states of the higher spin module. 

From the conformal field theory perspective, the calculations are rather standard, and we computed one and two-point functions of composite operators, with some of the details given in Appendix \ref{thermalcorrelators}. Given an ansatz for the currents as linear combinations of normal ordered composite operators, this allows the calculation of thermal traces involving the zero mode of the current and some simple low-weight operators such as the zero modes of normal ordered composite operators such as $(TT)_0$ and $(T'T')_0$.  

We then exploited the fact that the quantum Boussinesq charges are themselves zero modes of linear combinations of composite operators built out of $T$ and $W$. Thus, they act as linear operators at a given level of the higher spin module. This in turn allowed us to extract the excited state contribution to traces of the form $\langle {\bf I}_n\, (TT)_0\rangle $ or $\langle {\bf I}_n\, (T'T')_0\rangle $ by using oscillator methods and the commutation relations of the ${\cal W}_3$ algebra.  The calculations at the second excited level are done numerically, and we have provided the details in Appendix \ref{fixingJ9}. Having computed the trace by two completely independent means, a comparison leads to simple algebraic equations for the unknown coefficients appearing in the ansatz for the currents. Despite the numerical approach to the derivation of the equations, the result for these coefficients turns out to be simple rational numbers.

We note that we needed to use two sets of two-point functions to fix $J_{11}$ completely. We also had to compute the eigenvalues of the Boussinesq charges at the second excited level, which required some numerical interludes. The complications we have encountered in our analysis could be because we restricted our analysis to the insertion of operators built solely out of the stress tensor and its derivatives. It is likely that information regarding the undetermined constants could have been deduced more easily (perhaps using data from lower excited levels) if we inserted into the respective thermal correlators, operators involving the spin-3 current, such as $W_0$ or $(TW)_0$. It turns out, however, that taking these insertions into account necessitates the use of a generalized Zhu recursion (allowing for zero modes of multiple types of operators to appear in the recursion), which would require analysis of a different sort. In any case, it should be possible to extend our methods to compute any of the higher currents, provided we find simple enough (zero-mode) operators such that  (i) they have a non-vanishing correlator with the trace-free combinations appearing in the ansatz and (ii) we can compute the eigenvalues of the conserved charges in the higher excited levels. 

Since the form of the currents has been fixed, a natural next step is to calculate higher point thermal correlators involving the Boussinesq charges. From general arguments, these should be expressed as quasimodular differential operators acting on the character of the higher spin module. For the one-point functions of the quantum Boussinesq charges up to weight 10, this has been explicitly demonstrated in \cite{Ashok:2024zmw}. We have also checked that this continues to be true for two-point functions involving the low-weight charges. 

This naturally leads to the study of the generalized Gibbs ensemble and to the study of the partition function with fugacities turned on for all the mutually commuting charges \cite{Maloney:2018yrz, Dymarsky:2018lhf}. The (quasi-)modularity is crucial to extract the
high-temperature behaviour of the correlators (see \cite{Iles:2013jha, Iles:2014gra} for the high-temperature behaviour of the simplest $\langle {\bf I}_2^2\rangle = \langle W_0^2\rangle$ correlator), and the generalized partition function. This, in turn, should prove to be helpful in exploring the statistics of the Boussinesq charges and the Eigenstate Thermalization Hypothesis for (higher spin) conformal field theories \cite{Lashkari:2016vgj, Dymarsky:2019etq} in two dimensions. See the recent work \cite{Downing:2024nfb} in which the modular properties of the generalized Gibbs ensemble of quantum KdV charges have been studied. Another future goal would be to analyze the semi-classical limit of the spectrum of quantum Boussinesq charges and generalize the results of \cite{Dymarsky:2018iwx, Dymarsky:2022dhi} to the higher spin case. We hope to address these questions in the near future.

\section*{Acknowledgements}

We are grateful to Dileep P. Jatkar and Nemani V. Suryanarayana for helpful discussions, and we thank Jan Troost for his comments on an earlier version of this manuscript. 
This work received partial support from the INFN project SFT and the PRIN Project No. 2022ABPBEY, ``Understanding quantum field theory through its deformations''.

\begin{appendix}
    \section{Thermal correlators}
\label{thermalcorrelators}

We define the thermal correlator of operators ${\cal O}_i$ to be 
\be
\langle {\cal O}_1 \ldots {\cal O}_n\rangle = \Tr \, \left(q^{L_0-\frac{c}{24}}\, {\cal O}_1 \ldots {\cal O}_n\right)~,
\ee
where the elliptic nome is defined as $q = e^{-\beta}$, with $\beta$ being the inverse temperature. The trace is taken in the higher-spin module constructed over a highest weight state labeled by the eigenvalues $(\Delta_2, \Delta_3)$ of the commuting operators $(L_0, W_0)$. A generic state in the module will be of the form
\be
\mathcal{P}=\prod_{n>0}L_{-n}^{a_{n}}W_{-n}^{b_{n}}|\Delta_2, \Delta_3\rangle~.
\ee
We compute the thermal correlators using the Zhu recursion relation \cite{zhu:1990, zhu:1996} and their generalization given in \cite{ Gaberdiel:2012yb}. We omit most of the details of the calculation, as this has been reviewed in detail in our previous work \cite{Ashok:2024zmw}. In our ansatz for the currents, we have composite operators built out of $T(u)$,  $W(u)$ and their derivatives. These are conformally normal ordered; for an  operator made up of two fields $A_1(u_1)$ and $A_2(u_2)$, the normal ordered product $(A_1A_2)(u_1)$ is defined using the two-point thermal correlator \cite{Maloney:2018hdg}:
\be
\langle(A_1A_2)(u_1)\rangle=\frac{1}{2\pi i}\oint_{u_1}\frac{du_2}{u_2-u_1}\langle A_1(u_1)A_2(u_2)\rangle~.
\ee
For a $n$-point correlator involving $n$ such fields, we perform the normal ordering successively from right to left.  
    
\subsection{One-point functions of weight nine composites}
\label{listofthermalvevsI8} 
We present the thermal one-point functions of composite objects, which are relevant to compute the thermal one-point function of $I_8$. 
{\allowdisplaybreaks
    \begin{align*}
        \langle(T(T(TW)))(u)\rangle
        =&\partial^3 \langle W_0\rangle -\frac{5E_2}{4}  \partial^2 \langle W_0\rangle + \left(\frac{1}{480} (c+108) E_4+\frac{5 E_2^2}{12}\right) \partial \langle W_0\rangle \\ 
        &+\left(-\frac{(c+108) E_4 E_2}{1920}-\frac{13 (c+30) E_6}{12096}-\frac{5 E_2^3}{144} \right) \langle W_0\rangle \,,\\
        \langle(W(WW))(u)\rangle=&
        \langle W_0^3\rangle -\frac{E_2}{6}  \partial^2\langle W_0\rangle+\left(\frac{(5 c+166) E_4}{5760}+\frac{5 E_2^2}{72}\right) \partial\langle W_0\rangle\\
        &- 
        \left(\frac{\left(5 c^2+217 c+3930\right) E_6}{1451520}+\frac{(23 c+978) E_4 E_2}{69120}+\frac{E_2^3}{288}\right)\langle W_0\rangle \,,\\
        (2\pi)^2 \langle(T(WT^{\prime\prime}))(u)\rangle=&-\frac{E_{4}}{24}\partial\langle W_{0}\rangle+\left(\frac{(c+30) E_6}{3024}+\frac{E_2 E_4}{96}\right)\langle W_{0}\rangle ~,\\
           (2\pi)^2 \langle(T^{\prime}(T^{\prime}W))(u)\rangle=&\frac{E_{4}}{60}\partial\langle W_{0}\rangle -\left(\frac{(c+27) E_6}{3024}+\frac{E_2 E_4}{240}\right)\langle W_{0}\rangle~,\\
   (2\pi)^4 \langle(T^{\prime\prime}W^{\prime\prime})(u)\rangle=& -\frac{1}{84}E_{6}\langle W_{0}\rangle ~.
    \end{align*}
}
Using the above results, we can show the following combination is trace-free, in the sense that the thermal one-point function vanishes identically:
\be
\left \langle(T(WT''))(u)+\frac{5}{2}(T'(T'W))(u)-\frac{1}{24}(c+25)(2\pi)^2 (T''W'') (u)\right\rangle=0 ~.
\ee
Note that due to the reordering of the composite operator $(T'(T'W))$ as $(W(T'T'))$ in \cite{Ashok:2024zmw}, the trace-free combination is slightly different from the one listed in equation (4.24) of that reference. 

\subsection{Two-point functions involving weight nine composites}
\label{J9twopoint}
In this section, we provide the thermal two-point functions of composite objects, which are relevant to compute the thermal two-point function of $\langle {\bf A}_5I_8\rangle$. 

\begin{align*}
&(2\pi)^4 \left\langle\, {\bf A}_5 \, \int\, du\,  (T^{\prime\prime}W^{\prime\prime})(u)\right\rangle = \left(\frac{1}{30} E_2 E_4^2-\frac{169 E_4 E_6}{5040}\right) \partial \langle W_{0}\rangle\cr
   &+ \left(\frac{(210 c+1031) E_4^3}{110880}-\frac{(350 c+633) E_2 E_6 E_4}{110880}+\frac{(17695 c+66276) E_6^2}{13970880}-\frac{E_2^2 E_4^2}{120} \right) \langle W_{0}\rangle \,,\\
   \\
& (2\pi)^2 \left\langle  {\bf A}_5 \, \int\, du\,  (T^{\prime}(T^{\prime}W))(u)\right\rangle= \left(\frac{407 E_4^2}{25200}-\frac{E_2 E_6}{63} \right)\partial^2\langle W_0\rangle \\
&+\left(\frac{(560 c+7447) E_4^2 E_2}{302400}+\frac{(-1690 c-32139) E_4 E_6}{907200}+\frac{2}{189} E_6 E_2^2\right)\partial\langle W_0\rangle \\
&+\left( \frac{\left(-3500 c^2-106705 c-180486\right) E_4 E_6 E_2}{39916800}+\frac{\left(2100 c^2+73780 c+340341\right) E_4^3}{39916800}\right.\\
&\left. \hspace{1cm}+\frac{\left(17695 c^2+649881 c+2424492\right) E_6^2}{502951680}+\left(-\frac{c}{2160}-\frac{121}{16128}\right) E_4^2 E_2^2-\frac{1}{756} E_6 E_2^3\right)\langle W_0 \rangle \,,\\
\\
&(2\pi)^2   \left\langle\,  {\bf A}_5 \, \int\, du\,  (T(WT''))(u) \right\rangle = \left(\frac{5 E_2 E_6}{126}-\frac{407 E_4^2}{10080}\right)\partial^2 \langle W_0 \rangle\\
&+\left(\frac{(-392 c-3247) E_2 E_4^2}{120960}+E_6 \left(\frac{(1183 c+19464) E_4}{362880}-\frac{5 E_2^2}{189}\right)\right)\partial \langle W_0 \rangle \\
&+ \left( \frac{\left(-840 c^2-31774 c-177051\right) E_4^3}{15966720}+\frac{\left(-17695 c^2-671214 c-3194856\right) E_6^2}{502951680}\right.  \\
& \left.+E_6 \left(\frac{\left(1400 c^2+40039 c+64800\right) E_4 E_2}{15966720}+\frac{5 E_2^3}{1512}\right)+\left(\frac{7 c}{8640}+\frac{325}{32256}\right) E_2^2 E_4^2 \right)\langle W_0 \rangle \,, \\
\\
&    \left\langle\,  {\bf A}_5 \, \int\, du\,  (T(T(TW)))(u)\right\rangle =\frac{E_4}{60}\partial^4 \langle W_0\rangle-\left( \frac{(5 c+3024) E_6}{15120}-\frac{43 E_2 E_4}{240}\right)\partial^3 \langle W_0\rangle\\
&+\left(\frac{(-85 c-1044) E_6 E_2}{12096}+\frac{(1507 c+68906) E_4^2}{201600}-\frac{47}{192}  E_4 E_2^2\right)\partial^2\langle W_0 \rangle\\
&+\left( \frac{\left(280 c^2+30007 c+153316\right) E_4^2 E_2}{2419200}+\frac{145 E_4 E_2^3}{1728} \right. \\
&\left. \hspace{2cm}+E_6 \left(\frac{\left(-845 c^2-126754 c-2106252\right) E_4}{7257600}+\frac{(175 c+5112) E_2^2}{36288}\right)\right) \partial \langle W_0 \rangle\\
&+ \left(\frac{\left(-56 c^2-7207 c-85788\right) E_4^2 E_2^2}{1935360}+\frac{\left(54570 c^2+2088991 c+17218668\right) E_4^3}{319334400}\right.\\
&\left. +E_6 \left(\frac{\left(-80555 c^2-1987936 c-3875928\right) E_2 E_4}{319334400} 
-\frac{5 (53 c+1836) E_2^3}{435456}\right)  \right. \\
&\left. \hspace{4cm}+\frac{\left(222979 c^2+8161302 c+61790904\right) E_6^2}{2011806720}-\frac{49 E_4 E_2^4}{6912}
\right)\langle W_0 \rangle \,,\\
\\
&\left    \langle\,  {\bf A}_5 \, \int\, du\,  (W(WW))(u)\right\rangle =\frac{E_4}{60}\partial\langle W_0^3\rangle+\left(\frac{E_2E_4}{5} -\frac{(5 c+3024) E_6}{15120}\right)\langle W_0^3\rangle\\
&+\frac{E_2E_4}{360}\partial^3 \langle W_0\rangle+ \left( \frac{(5 c-1476) E_6 E_2}{90720}+\frac{(35 c+121722) E_4^2}{2419200}-\frac{7}{216} E_4 E_2^2\right)\partial^2\langle W_0 \rangle \\
&+\left(E_6 \left(\frac{\left(-10 c^2-53689 c-1274918\right) E_4}{29030400}+\frac{(4176-5 c) E_2^2}{217728}\right)\right.\\
&\left. \hspace{5.5cm}+\frac{(53179 c+301162) E_4^2 E_2}{29030400}+\frac{727 E_4 E_2^3}{51840} \right)\partial \langle W_0 \rangle \\
&+ \left(\frac{\left(46460 c^2+2074265 c+29875806\right) E_4^3}{3832012800}+\frac{(-168469 c-2879910) E_4^2 E_2^2}{348364800}\right.\\
&\left. +E_6 \left(\frac{\left(-238735 c^2-4379196 c+10311732\right) E_4 E_2}{11496038400}+\frac{(5 c-14976) E_2^3}{4354560}\right)\right.\\
&\left. \hspace{3cm}+\frac{\left(275 c^3+2124955 c^2+79379178 c+898468200\right) E_6^2}{241416806400}-\frac{49 E_4 E_2^4}{69120} \right) \langle W_0 \rangle ~.
\end{align*}
From these explicit results, one can check that
\be \label{A5withI8L2}
\begin{aligned}
    \Big\langle 
    {\bf A}_5\, 
   \Big((T(WT''))+\frac{5}{2}(T'(T'W)) & -\frac{1}{24}(c+25)(2\pi)^2 (T''W'') \Big)(u)
    \Big\rangle \cr
    &= \frac{c+2}{18480}(5E_4^3+7E_6^2-12E_2E_4E_6)
   \langle W_0 \rangle\cr
    &= 216(c+2)\Delta_3\, q^{\Delta_2-\frac{c}{24}}\, ( q^2 + 58q^3  + \ldots)~ .
\end{aligned}
\ee
Thus, there will be a $\gamma$-dependent contribution at $O(q^2)$ to the 
two-point function of ${\bf A}_5$ and ${\bf I}_8$. By combining the  thermal two-point functions of ${\bf A}_5$ and the composite operators appearing in the ansatz for $J_9$, the $O(q^2)$ contribution to the two-point function is given by:  
\begin{align}
\label{A5I82point}
    \langle {\bf A}_5\, {\bf I}_8 \rangle_{L=2} &= \frac{529}{12} (\Delta _2^4 \Delta _3+3\Delta_3^3\Delta_2)+\left(\frac{144337}{48}-\frac{4583 c}{3024}\right) \Delta _2^3 \Delta _3+\left(\frac{24161 c}{2016}+\frac{17329}{2}\right) \Delta _3^3\cr
    &+\left(-\frac{3263 c^2}{12096}+\frac{45792023 c}{120960}+\frac{2285741}{64}\right) \Delta _2^2 \Delta _3\\
    &+\left(\frac{10229 c^3}{580608}+\frac{71822327 c^2}{2903040}+\frac{52879403 c}{30240}+\frac{152645845}{2304}\right) \Delta_2\Delta _3\cr
    &+\Delta _3 \Big(-\frac{24161 c^4}{83607552}+\frac{87748349 c^3}{1463132160}+\frac{3402082927 c^2}{81285120}+ \frac{4070183401}{5806080} c+\frac{24989101}{1152}\cr
    & \hspace{11cm}+216\gamma^{(0)}(c+2)\Big)~. \nonumber
\end{align}
This result will be used in Appendix \ref{fixingJ9}, and compared with the result for the two-point function calculated using the operator formalism. 

To fix the form of ${\bf I}_{10}$, similar results for the one and two-point correlators have been computed for the individual composite operators appearing in the $J_{!1}$ ansatz. We do not present the analytical form of the results as they are quite cumbersome. However, we shall present numerical results for these in the next section for specific choices of $(c, \Delta_2, \Delta_3)$.

\section{Fixing the Currents: Numerical Details}
\label{fixingJ9}

In this section we collect the details of the calculations that fix the higher Boussinesq currents. 

\subsection{\texorpdfstring{$J_{9}$}{J9}}

We begin with fixing the undetermined constant in $J_9$. 
\begin{itemize}
     \item First, we choose a random set of values for the central charge and conformal dimensions of the module under consideration. This can be seen in the first column of Table \ref{table:1}. 
     
\item Secondly, we solve for the algebraic constraints in \eqref{monoconst2}. We note that we have solved the equations numerically up to 20 digits of precision. We find exactly five solutions that correspond to the eigenvectors of the quantum Boussinesq charges at the second excited level. Substituting these values into the eigenvalue of ${\bf I}_8$ computed in the previous section gives the five eigenvalues at the second excited level. These are listed in the second column of Table \ref{table:1} for ${\bf I}_8$. 
     \item The next step computes the trace in the operator formalism. To do so, we first compute the five eigenvalues of $W_0= {\bf I}_2$ using \eqref{I2level2}. We then diagonalize the matrix \eqref{W0inlevel2basis} to compare with this list and find the R-matrix that implements this diagonalization. Since the charges all mutually commute, this gives us the R-matrix that also diagonalizies ${\bf I}_8$. Substituting this into  the formula for the trace (see \eqref{levelLtrace}):
     \be
\langle {\bf I}_8\, {\bf A}\rangle_{{\cal H}_{2}}  =  q^{\Delta_2+2}\,  \Tr\left( A\, R\, I_{n}^{(L)}\, (R^{-1})\right)~,
\ee
we find the entries in the third column of Table \ref{table:1}. 
\item To obtain the final column, we simply substitute the values of $(c,\Delta_2, \Delta_3)$ into the $O(q^2)$ contribution to the trace computed using the thermal correlators, and presented in \eqref{A5I82point}.
\begin{table}
\begin{center}
{\scriptsize
\begin{tabular}{ |c|c|c|c| } 
 \hline
 $\big(c,\Delta_{2},\Delta_{3}\big)$ 
 & Eigenvalues at level 2: ${I}_{8,i}^{(2)}$ 
 & $\langle {\bf A}_{5} {\bf I}_{8}\rangle_{L=2}$ (osc.) 
 & $\langle {\bf A}_{5} {\bf I}_{8}\rangle_{L=2}$ (therm.) \\
 \hline
 \hline
& -8245.7932795505   &  &  \\ 
& -663.54732556260  &  & 2715.29003975634 $\gamma^{(0)}$   \\ 
$\big(-10,\frac{20}{9},-\frac{10\sqrt{2}}{9}\big)$&  515.28751158324   &  -542700.22989155        &-540663.76236173 \\
 &  43.77545199884   &          &  \\ 
 & -483.08161460450   &          &  \\ 
 \hline

\end{tabular}\label{table:1}
\caption{Numerical results that lead to the determination of $\gamma^{(0)}$. We first choose particular values of $(c, \Delta_2, \Delta_3)$. This allows us to solve for the $(a_i, w_i)$ numerically. We find five distinct solutions. Using these, we find the five distinct eigenvalues of ${\bf I}_8$ at the second excited level and the trace with ${\bf A}_5$ inserted. We compare this with the result for the trace calculated using the thermal two-point functions, and this allows us to fix $\gamma^{(0)}$ uniquely.}
}
\end{center}
\end{table}

\item Equating the entries in the third and fourth column, we find $\gamma^{(0)}$ to be given by:
\be 
\gamma^{(0)} = - \frac34~.
\ee 
We obtain the same value of $\gamma^{(0)}$ by choosing various other choices of $(c, \Delta_2, \Delta_3)$. 
\end{itemize}

\subsection{\texorpdfstring{$J_{11}$}{J11}}

We now repeat these steps for the case of ${\bf I}_{10}$. Recall that there are six undetermined constants in the expression for $J_{11}$ in \eqref{J11afterclassical}. We first consider the two-point function of the charge with ${\bf A}_5$. At level 1, we find that no matter what numerical values for $(c,\Delta_{2},\Delta_{3})$ are chosen, we obtain a single equation, given by
\be 
\label{level1eqn}
8\delta_3^{(0)}-28 \delta_4^{(0)} +1 = 0~.
\ee
Moving on to level 2, we repeat the steps outlined earlier; we obtain the results in Table \ref{table:2}.
\begin{table}
\begin{center}
{\scriptsize
\begin{tabular}{ |c|c|c|c| } 
 \hline
 $\big(c,\Delta_{2},\Delta_{3}\big)$ 
 & Eigenvalues at level 2: ${I}_{10,i}^{(2)}$ 
 & $\langle {\bf A}_{5} {\bf I}_{10}\rangle_{L=2}$ (osc.) 
 & $\langle {\bf A}_{5} {\bf I}_{10}\rangle_{L=2}$ (therm.) \\
 \hline
 \hline
 &3533.5725819+ $i$2.5725964257922$\times 10^7$  & &-8.42712458599599174$\times 10^6$\\
  &3533.5725819- $i$2.5725964257922$\times 10^7$  & &-237.3472222222222222 $\delta^{(0)}_{1}$\\
$\left(-\frac{671}{2},-\frac{1009}{72},-\frac{1}{288}\right)$   &19.40887956686+ $i$794.15593683822  &-8.547506010$\times 10^6$ &+79629.9930555555556 $\delta^{(1)}_{1}$\\
&19.40887956686- $i$794.15593683822  & &-494757.394811820083 $\delta^{(0)}_{4}$\\
&64.743553579092 & &+141553.1584382978014 $\delta^{(0)}_{3}$\\
\hline
&131.773179900298+ $i$2561.55911926717  & &158427.841348086225\\
&131.773179900298- $i$2561.55911926717  & &-2386.41125022697556 $\delta^{(0)}_{1}$\\
$\left(-\frac{374}{5},-\frac{116}{45},-\frac{16}{45 \sqrt{5}}\right)$&-15.1220957985725 &-82937.229195105 &+178503.561516977772 $\delta^{(1)}_{1}$ \\
&-3.00689419219823+ $i$2.42552596973312  & &-116476.3098370625835 $\delta^{(0)}_{4}$\\
&-3.00689419219823- $i$2.42552596973312 i & &+39746.9342833526784 $\delta^{(0)}_{3}$\\
\hline
&431.226110418+ $i$338276.69028687  & &-420167.815576069929\\
&431.226110418- $i$338276.69028687  & &-289.520943185825292 $\delta^{(0)}_{1}$\\
$\left(-145,-\frac{217}{36},-\frac{1}{72 \sqrt{2}}\right)$&-8.25576901577- $i$645.02009814402  &-478157.823246 &+41980.5367619446673 $\delta^{(1)}_{1}$\\
&-8.25576901577+ $i$645.02009814402  & &-97403.7929087827859$\delta^{(0)}_{4}$\\ 
&9.9242777690861 & &+28419.3456945738868 $\delta^{(0)}_{3}$\\
    \hline
    &131.773179900298+ $i$2561.55911926717  & &156415.623479858187\\
    &131.773179900298- $i$2561.55911926717  & &-2074.75306456300487 $\delta^{(0)}_{1}$\\
$\left(-\frac{374}{5},-\frac{101}{45},-\frac{14}{45 \sqrt{5}}\right)$    &-15.1220957985725 &-54298.791240984 &+155191.529229312764 $\delta^{(1)}_{1}$\\
    &-3.00689419219823+ $i$2.42552596973312  & &-78260.5005359722806 $\delta^{(0)}_{3}$\\
    &-3.00689419219823- $i$2.42552596973312  & &-78260.5005359722806 $\delta^{(0)}_{4}$\\
    \hline
\end{tabular}\label{table:2}
    \caption{Numerical results required to find constants involved in $\mathbf{I}_{10}$. As it can be seen explicitly the thermal correlator of $\mathbf{I}_{10}$ with $\bf{A}_{5}$ involves only four undetermined constants ($\delta^{(0)}_{1}$,$\delta^{(1)}_{1}$,$\delta^{(0)}_{3}$,$\delta^{(0)}_{4}$). These constants can be fixed from four numerical choices of $(c, \Delta_2, \Delta_3)$.}
    }
\end{center}
\end{table}

We note that when we equate the entries of the third and fourth columns for each of the rows, we find four linearly independent equations in four variables, which we solve to find
\be 
\label{betasigmadelta}
\begin{aligned}
    \delta_1^{(1)} &=~-\frac{253}{192},  \quad \delta_1^{(0)} = -\frac{879}{64}\,,  \\
    \delta_3^{(0)} &=-\frac{43}{8} ~, \quad \delta_4^{(0)} = -\frac{3}{2}~.
\end{aligned}
\ee
We can immediately check that the values of $\delta_3^{(0)}$ and $\delta_4^{(0)}$ satisfy the equation in \eqref{level1eqn}. Furthermore, we have done analogous calculations for other values of $(c,\Delta_2, \Delta_3)$ and found exactly the same results. This leaves two parameters to fix. To fix these, we compute the two-point function of ${\bf I}_{10}$ with ${\bf A}_3$, and the results are shown in Table \ref{table:3}. 

\refstepcounter{subsection}
\begin{table}[H]
\begin{center}
{\scriptsize
\begin{tabular}{ |c|c|c|c| }  
 \hline
 $\big(c,\Delta_{2},\Delta_{3}\big)$ 
 & Eigenvalues at level 2: ${I}_{10}^{(2)}$ 
 & $\langle {\bf A}_{3} {\bf I}_{10}\rangle_{L=2}$ (osc.) 
 & $\langle {\bf A}_{3} {\bf I}_{10}\rangle_{L=2}$ (therm.) \\
 \hline
 \hline
&-13.08889619- $i$216.69461435  & &28809.94074427 - 333.33333333   $\delta^{(0)}_{1}$\\
&-13.08889619+ $i$216.69461435  & &17333.33333333 $\delta^{(1)}_{1}$\\
$\left(-52,-\frac{31}{18},-\frac{5}{36}\right)$&-45.72705707 &891.8628026497&+1500 $\delta^{(0)}_{2}$-78000 $\delta^{(1)}_{2}$\\
&-0.11270168- $i$1.21762082  & &3497.40131077 $\delta^{(0)}_{3}$\\
&-0.11270168+ $i$1.21762082  & &-8470.07125438 $\delta^{(0)}_{4}$\\
\hline
&2570.33616417+ $i$273006.46983069  & &-471.87592531$\delta^{(0)}_{1}$+68422.00917021$\delta^{(1)}_{1}$\\
&2570.33616417- $i$273006.46983069  & &+2123.44166390$\delta^{(0)}_{2}$-495890.02280547\\
$\left(-145,-\frac{199}{36},-\frac{7}{72 \sqrt{2}}\right)$&-37.18463907 -$i$262.65391607   &-608361.8110576 &+43653.74702321$\delta^{(0)}_{3}$\\
&-37.18463907+ $i$262.65391607  & &-155678.35462377$\delta^{(0)}_{4}$\\
&23.58335121  & &-307899.04126596$\delta^{(1)}_{2}$\\

\hline
 \end{tabular}
    \caption{Numerical results required to find ($\delta^{(0)}_{2}$,$\delta^{(1)}_{2}$). These constants can be fixed from the thermal correlator of $\mathbf{I}_{10}$ with $\bf{A}_{3}$ with two different numerical choices of $(c, \Delta_2, \Delta_3)$.}
    }
    \label{table:3}
\end{center}
\end{table}
Equating the third and fourth columns for each row and substituting the values in \eqref{betasigmadelta}, we obtain
\be
 \delta_2^{(1)} =\frac{11}{96} ~, \quad \delta_2^{(0)} =\frac{43}{12}~.
\ee
We have checked that the same results for the coefficients are obtained for a variety of values chosen for $(c, \Delta_2, \Delta_3)$. 

\label{excitedstateEV}
\end{appendix}

\bibliographystyle{JHEP}

\end{document}